\documentclass{aa}
\usepackage{txfonts}
\usepackage{graphicx}
\begin{document}

\title{The empirical upper limit for mass loss of cool main sequence stars}

\author{Anna Lednicka \and  K. St\c epie\' n}

\offprints{K. St\c epie\'n}
\institute{Warsaw University Observatory, Al. Ujazdowskie 4, 00-478
Warszawa, Poland\\
email: lednicka@astrouw.edu.pl, kst@astrouw.edu.pl}
\date{Received ...;accepted ...}

\abstract 
{The knowledge of mass loss rates due to thermal winds in cool
dwarfs is of crucial importance for modeling evolution of physical
parameters of main sequence single and binary stars. Very few, sometimes
contradictory, measurements of such mass loss rates exist up to now.}  
{We
present a new, independent method of measuring an amount of mass lost by a
star during its past life.} 
{It is based on the comparison of the present
mass distribution of solar type stars in an open cluster with the
calculated distribution under an assumption that stars with masses lower
than $M_{\mathrm{lim}}$ have lost an amount of mass equal to $\Delta
M$. The actual value of $\Delta M$ or its upper limit is found from the
best fit.} 
{Analysis of four clusters: Pleiades, NGC 6996, Hyades and
Praesepe gave upper limits for $\Delta M$ in three of them and the
inconclusive result for Pleiades. The most restrictive limit was obtained
for Praesepe indicating that the average mass loss rate of cool dwarfs in
this cluster was lower than $6\times 10^{-11} M_{\sun}$/yr. With more
accurate mass determinations of the solar type members of { selected
open clusters}, including those of spectral type K, the method will provide
more stringent limits for mass loss of cool dwarfs.}
{}


\maketitle

\section{Introduction}
Chromospheric-coronal activity is ubiquitous in cool stars possessing
subphotospheric convection zones. Hot coronas result in existence of
thermal winds carrying away stellar mass and angular momentum. The precise
knowledge of the mass loss rate (MLR) during the stellar life is of crucial
importance for modeling evolution of mass, luminosity and rotation of
single stars and of orbital parameters of close binary stars. Mass loss of
a central star of a planetary system may also influence evolution of
physical parameters of orbiting planets.  Proximity of the Sun made
possible direct measurements of solar wind with the use of interplanetary
probes. The average velocity of the wind near the Earth orbit is of the
order of 400 km s$^{-1}$ and the average MLR is $2\times 10^{-14}
M_{\sun}$/yr with uncertainty of some 50 \% (Feldman et al. 1977). The
expected MLRs for other cool dwarfs may differ from that at most by a
couple of orders of magnitude but even the highest MLRs of such stars are
far lower than those occurring in hot, early type stars or red giants. The
present observational capabilities are still insufficient to obtain the
data for other dwarfs with comparable to the solar case accuracy. For many
years only indirect methods of MLR estimate of such stars
existed. Analyzing the infrared and radio observations of M dwarfs Mullan
et al.  (1992) obtained a value of $10^{-10} M_{\sun}$/yr as an upper limit
for MLR of these stars. Later, van den Oord \& Doyle (1997) revised that
limit down to $10^{-12} M_{\sun}$/yr. A similar value was obtained by Lim
\& White (1996). Wargelin \& Drake (2002) obtained an upper limit of
$3\times 10^{-13} M_{\sun}$/yr for {\it Proxima} Cen. In the recent years
Wood and coworkers published a number of papers in which they analyzed
profiles of L$_{\alpha}$ arising at the collisional front between stellar
wind and interstellar matter. By modeling the collision process the authors
were able to measure MLRs for a dozen stars (Wood et al. 2002, 2005). The
measured rates extend from $2\times 10^{-12} M_{\sun}$/yr for 70 Oph down
to $3\times 10^{-15} M_{\sun}$/yr for DK UMa. The authors derived a
relation between MLR and age (or X-ray flux). Their latest results give
$\dot M \propto t^{-2.33}$ for $t \ge 0.7$ Gyr but observations of younger,
more active stars suggest a sudden drop of MLR from about $2\times
10^{-12}$ to $\sim 10^{-13} M_{\sun}$/yr, i.\, e. by more than an order of
magnitude. This unexpected result is based on scanty data and needs
confirmation.  A different, indirect method of estimation an amount of mass
lost by the Sun during its past life was suggested by Sackmann \& Boothroyd
(2003). Based on the geological evidence that the terrestrial atmosphere
was warm in early life of the Solar System and liquid water was present on
Mars, they calculated the minimum initial mass of the Sun needed to produce
enough luminosity, { and reconstructed time variation of the solar MLR
so adjusted that the favorable thermal conditions on the Earth and Mars
have been kept as the Sun aged. The optimum conditions were reproduced} for
the initial mass of 1.07 $M_{\sun}$ and MLR decreasing exponentially with
time.

Here we { present} another method of measuring the total amount of mass
lost during the stellar life. It is based on the analysis of the stellar
mass distribution in an open cluster. The next section describes details of
the { suggested} method. The last section contains discussion of the
results and conclusions.

\section{Method}

\subsection{General considerations}

The total energy flux carried away with the stellar wind,
$F_{\mathrm{tot}}$, consists of three components (Holzer 1987)

\begin{equation}
F_{\mathrm{tot}} = F_{\mathrm{grav}} + F_{\infty} +F_{\mathrm{rad}}\,,
\end{equation}

where $F_{\mathrm{grav}}$ is the energy flux needed to carry the wind
matter out of the potential well of the star, $F_{\infty}$ is the kinetic
energy flux of the wind in infinity and $F_{\mathrm{rad}}$ is the flux
radiated away by the wind. Based on observations of the solar wind
$F_{\mathrm{rad}}$ can be neglected. We thus obtain

\begin{equation}
F_{\mathrm{tot}} = \frac{1}{2}\dot M(v^2_{\mathrm{esc}} + v^2_{\infty})
  \approx \dot Mv^2_{\mathrm{esc}}\,,
\end{equation}

where $v_{\mathrm{esc}}$ is the escape velocity from the stellar surface
and $v_{\infty}$ is the wind velocity in infinity. The last approximate
equality results from the assumption that both velocities are of the same
order. With $v_{\mathrm{esc}} = (2GM/R)^{1/2}$, where $G$ is gravitational
constant, and  $M$ and $R$ are 
stellar mass and radius, respectively, we have

\begin{equation}
F_{\mathrm{tot}} = \frac{2GM\dot M}{R}\,.
\end{equation}

The ultimate source of energy for thermal winds in cool stars is stellar
luminosity which drives convection and, via a chain of physical processes,
makes a magnetized wind to blow. Rotational energy, albeit necessary for
magnetic field generation (dynamo does not work in nonrotating stars), is
of secondary importance. If so, the total energy flux of the wind can be
expressed as a fraction $kL$ of stellar luminosity

\begin{equation}
\dot M = 1.5\times 10^{-8}\frac{kLR}{M}\,
\end{equation}

{ where $M$, $R$ and $L$ are in solar units and $\dot M$ is expressed in solar mass
per year. The observed solar value of MLR gives} $k_{\sun} = 1.3\times
10^{-6}$. For comparison, Kudritzki \& Reimers (1978) derived an empirical
relation for M-type giants

\begin{equation}
\dot M = 5.5\times 10^{-13}\frac{LR}{M}\,,
\end{equation}

{ from which $k_{\mathrm{MG}} = 3.7\times 10^{-5}$. Surprisingly,
although MLRs of M-giants are some 7 orders of magnitude higher than the
solar value, the efficiency factor $k_{\mathrm{MG}}$, describing the
transformation of the red giant radiation into wind energy, is only about
30 times higher than $k_{\sun}$}.

How high values can MLR reach in the most active cool dwarfs?  Stars with
X-ray emission at the saturated level have $L_x \approx 10^{-3}L$ and if we
apply the same efficiency factor to MLR (i.\thinspace e. $k = 10^{-3}$) we
obtain an upper limit of $1.5\times 10^{-11} M_{\sun}$/yr for a one solar mass
star. The upper limit drops, however, sharply for lower mass stars due to a
fast decrease of their luminosity and reaches a value of $7\times 10^{-13}
M_{\sun}$/yr for a 0.4 solar mass star. (Note that, according to the
results of Wood et al. 2005, the proportionality between MLR and X-ray
luminosity breaks down for the most active stars). This brief discussion
indicates that the expected MLR of active solar type stars should not
considerably exceed $\sim 10^{-11} M_{\sun}$/yr and yet it may be significantly
lower if the results of Wood et al. (2005) are confirmed.

\subsection{Method of the mass loss estimate}

Initial mass distribution of a stellar cluster is well described by the
Salpeter function (Salpeter 1955). Recent discussion of this distribution
by Kroupa (2002) showed, as the author stresses, that the initial mass
function of different populations of stars shows an extraordinary
uniformity and is well described by the Salpeter function, contrary to the
simple minded reasoning that it should vary with star-forming
conditions. The Salpeter mass distribution is given by

\begin{equation}
\frac{\mathrm{d}N}{\mathrm{d}M} \propto M^{-\alpha}\,, 
\end{equation}

where $\mathrm{d}N$ is a number of stars with masses between $M$ and $M +
  \mathrm{d}M$. { Values of $\alpha$ determined for individual star
  aggregates in the intermediate mass range concentrate around the original
  Salpeter value $\alpha = 2.35$ with a scatter expected from statistics
  (Elmegreen 2001).}
 
\begin{figure}
\resizebox{\hsize}{!}{\includegraphics{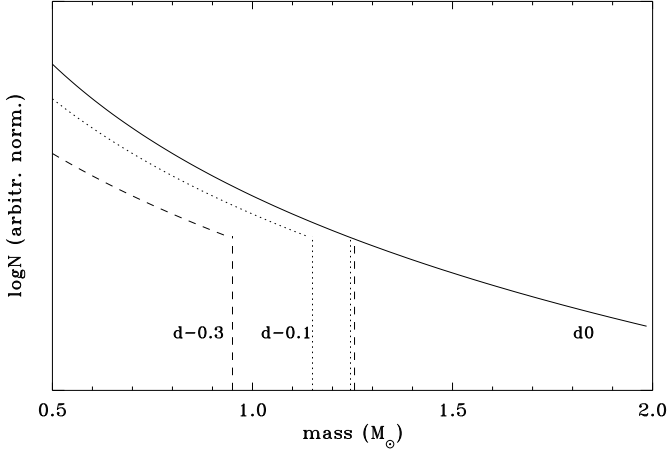}}
\caption{Expected differences among stellar mass distributions in a
cluster, resulting from the assumed mass loss: solid line, labelled d0 - 
no mass loss;
dotted line, labelled d-0.1 - mass loss of 0.1 $M_{\sun}$ in stars with $M <
1.25 M_{\sun}$; broken line, labelled d-0.3 - mass loss of 0.3 $M_{\sun}$ in
the same stars. All three distributions have a common part for stars with
$M \ge 1.25 M_{\sun}$ { (the vertical dotted and dashed lines at $M =
  1.25 M_{\sun}$ are slightly displaced from their exact positions for
  better visibility)}. Note a gap in two latter distributions, broadening with
the increased mass loss.} 
\label{fig1}
\end{figure}

The mass distribution evolves with the cluster age due to different
dynamical effects, like evaporation of low-mass stars and stellar evolution
affecting mostly high-mass stars. The mass distribution of solar
type stars (with masses, say, between 0.5 and 2 $M_{\sun}$) is expected to
be least affected by the above mentioned effects. It will, however, evolve
due to mass loss via thermal wind from stars with subphotospheric
convection zones. Let us assume that the mass distribution of
solar type stars with $M > M_{\mathrm{lim}}$ has not changed during the
past cluster life but stars with $M \le M_{\mathrm{lim}}$ have lost an
amount of mass equal to $\Delta M$. The assumption is justified by the
sharp rise of coronal activity at mid-F spectral type and, equally sharp,
drop of stellar rotation velocity around the same type, indicating a sudden
appearance of magnetized winds over a narrow spectral range (Schmitt 1997;
Gray 2005). Let $\Delta M$ be independent of mass. A gap in
the mass distribution then occurs just below $M_{\mathrm{lim}}$ with a
width of $\Delta M$ and, at the same time, the distribution function for
lower mass stars will run lower than the initial function. Fig~\ref{fig1}
shows the initial (Salpeter) mass distribution and two distributions
modified by mass loss of 0.1 and 0.3 $M_{\sun}$ when $M_{\mathrm{lim}} =
1.25 M_{\sun}$. The second value of $\Delta M$ is probably unrealistically
high but it is shown here to emphasize the expected distribution
modifications.

There are indications that the bulk of mass loss occurs when stars are
young and active i.\, e. younger than about 1 Gyr. The time-scale of
spin-down of young, rapidly rotating stars is about $1-2\times 10^8$ years
(St\c epie\'n 1988) which means that after several hundred Myr the activity
drops considerably. The observations of stellar X-ray activity by G\" udel,
Guinan \& Skinner (1997) show, indeed, that the activity decreases by about
1.5 order of magnitude during the first Gyr of stellar life. These results
indicate that the optimum cluster age for the analysis of $\Delta M$ is
from several hundred Myr, up to about 1 Gyr. { Stars in very young
clusters have not had enough time to lose an appreciable amount of mass
whereas $\Delta M$ in old clusters is expected to increment negligibly with
age due
to a decrease of MLR down to the solar value or even beyond. At the
same time, mass distribution of an old cluster becomes notably perturbed by
other effects, like evaporation of low mass stars and evolution of massive
stars.} With $\Delta M$ known, the {\em average} MLR can be estimated:
$\dot M_{\mathrm{av}} = \Delta M/T$, where $T$ is the cluster age.

{ Unfortunately, accurate mass values of solar mass members are
  presently known only
for very few clusters in the optimum age range.  We selected four clusters
for the analysis: Hyades, Praesepe, Pleiades and NGC 6996. Masses of Hiads
were { determined} by Perryman et al. (1998). Later analysis by de Bruijne,
Hoogerwerf \& de Zeeuw (2001) confirmed these values.} Perryman et
al. (1998) give masses for 218 stars. After rejecting all giants, variables
and stars marked by de Bruijne et al. (2001) as nonmembers, 133 dwarfs were
left. Data on stellar masses of Praesepe and Pleiades were taken from
Raboud \& Mermilliod (1998a, b). The authors selected 185 members of
Praesepe and 270 members of Pleiades and determined their masses from
colors using isochrones with metallicity Z = 0.02. { Masses of known binary
components were determined as described by Raboud \& Mermilliod (1998a)}. 
No published masses for
members of NGC 6996 exist so we computed them using $UBV$ photometry of
Villanova et al. (2004). The following values of the parameters, needed to
convert the observations into absolute photometry, were adopted: distance
moduli 9.4 mag, $E(B-V) =0.54$ and $A_V = 3E(B-V)$. Bolometric corrections
were applied following Bessel, Castelli \& Plez (1998). Individual masses
were calculated from the relation (Lang 1992)

\begin{equation}
\log\left(\frac{M}{M_{\sun}}\right) = 0.263\log\left(\frac{L}{L_{\sun}}\right) -
0.021\,.
\end{equation}

Photometric errors result in $\pm 0.05 M_{\sun}$ random mass errors. The
errors of absolute values of stellar masses are likely to be larger, due to
{ probable} systematic errors, but the uncertainties of the {\em
relative} mass determinations should be close to those resulting from
photometry. { Errors of mass determination of single stars in the other
analyzed clusters are assumed to be of the same order i.\thinspace e.  $\pm
0.05 M_{\sun}$. In case of resolved binaries with no spectroscopic orbits
expected errors are several times larger. Unresolved binaries present
another problem. They enter the analysis as single objects instead of two
stars, producing thus a bias in the observed mass
distribution. Fortunately, the commonly employed methods of detecting
binaries are most sensitive to systems with similar components, i.e. when
mass ratio $q$ is close to 1. As a result, frequency of known binaries
among solar type stars in the investigated clusters is as high as, or
higher than, among nearby field stars for which binary searches are
believed to be nearly complete (Halbwachs et al. 2004). This indicates that
in well studied clusters the percentage of missing binaries with {\em both}
component masses from the considered mass range [0.5,2.0] $M_{\sun}$ is
low, except, possibly, in the cluster central regions where crowding of
stellar images makes analysis more difficult. A stronger bias is expected
among low $q$ binaries with a solar type primaries and very low mass
secondaries. Only primaries enter the statistics and secondaries are
missing, which produces a deficit in the lowest mass tail of mass
distribution (Kroupa 1995). Such a deficit does not matter in our case as
long as we restrict the analysis to solar type stars only.}

We assume conservatively that two mass distributions differing in $\Delta
M$ by less than $0.05 M_{\sun}$ are statistically indistinguishable, so if
for example the original Salpeter function correctly describes the observed
distribution of a given cluster, we adopt 0.05 $M_{\sun}$ as an upper limit
for mass loss of its solar type members.

The observed mass distribution in each analyzed cluster was binned in 0.1
$M_{\sun}$ wide intervals. { Multiple star components with known masses
were treated as single stars.} The calculated distribution, corresponding
to a given value of mass loss, was calibrated against the mass interval
$1.25 - 2.0 M_{\sun}$ and its part corresponding to $M \le 1.25 M_{\sun}$
was compared to the observations.  The Kolmogorov-Smirnov test was applied
to verify a null hypothesis that the calculated distribution is identical
with the observed one in this mass interval. { The test uses the
statistic $\lambda$ (sometimes denoted by $D_n$) describing the maximum
difference between the empirical and the given (theoretical) cumulative
distribution functions}.  We adopted the significance level of 5\% which
corresponds to the statistic $\lambda \le \lambda_{\rm{lim}} = 1.36$. The
procedure was applied to the Salpeter function and the distributions with a
few values of $\Delta M$. The results are shown in Table~\ref{fit}.

\begin{table}
\caption[]{Results of a comparison of the calculated and observed mass
  distributions. The compared distributions are identical at the
  significance level of 5\% if statistic $\lambda \le \lambda_{\rm{lim}} = 
1.36$.}
\label{fit}
\begin{tabular}{llccllc} 
\hline

Cluster & $\Delta M$ & $\lambda$ && Cluster & $\Delta M$ & $\lambda$ \\
\hline
Pleiades & 0 & 2.18 && Hyades & 0 & 0.58 \\
         & 0.1 & 2.96 &&       & 0.1 & 0.93 \\
         & 0.2 & 3.81 &&       & 0.2 & 1.42 \\
\hline
NGC 6996 & 0 & 0.55 && Praesepe & 0 & 1.08 \\
         & 0.1 & 1.05 &&         & 0.1 & 1.75 \\
         & 0.2 & 1.58 &&         & 0.2 & 2.62 \\
\hline
\end{tabular}
\end{table}

\section{Results}

\subsection{Comparison of the calculated mass distributions to observations}

\begin{figure}
\resizebox{\hsize}{!}{\includegraphics{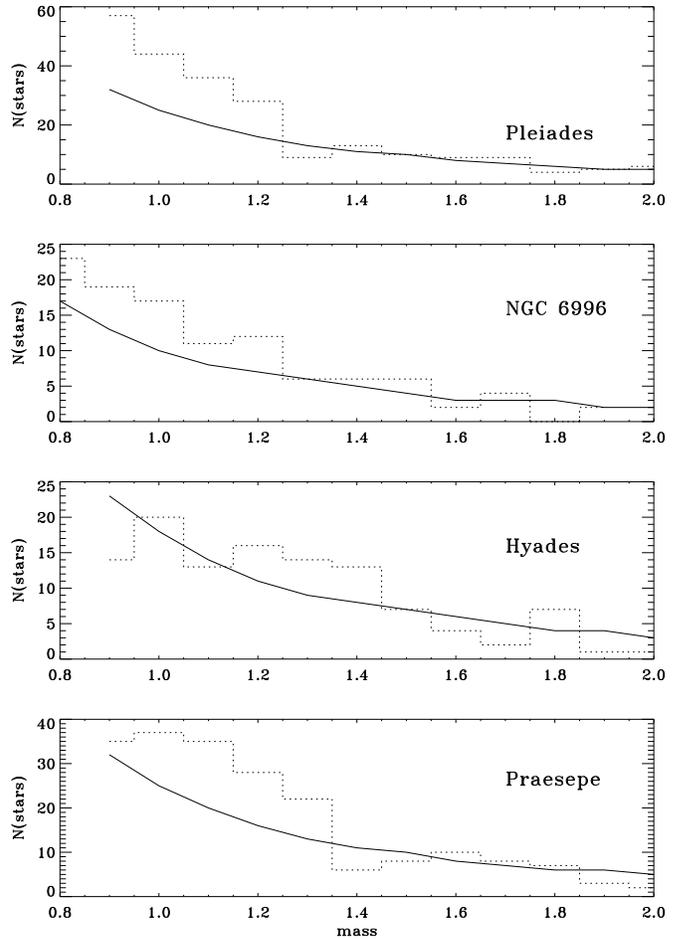}}
\caption{The observed mass distributions in three clusters with the
  Salpeter functions over plotted. The Salpeter function is normalized in
  each case to the distribution of stars with masses from the interval
  1.25-2.0 $M_{\sun}$.}
\label{fig2}
\end{figure}

{ Assuming that solar type members of a given cluster have lost $\Delta
M_{\rm{cl}}$ mass over their past life, we expect the function
$\lambda(\Delta M)$ to reach a minimum $\lambda = \lambda_{\rm{min}}$ at
$\Delta M = \Delta M_{\rm{cl}}$. If the condition $\lambda_{\rm{min}} \le
\lambda_{\rm{lim}}$ is fulfilled we assume that $\Delta M_{\rm{cl}}$ is the
most probable value of the amount of mass lost by the cluster members
although any value of $\Delta M$ from the neighbourhood of $\Delta
M_{\rm{cl}}$ is { also possible} as long as $\lambda(\Delta M) \le
\lambda_{\rm{lim}}$. The latter condition determines the uncertainty of the
obtained mass loss. Its rigid application may, however, result in an
underestimation of actual uncertainty of $\Delta M_{\rm{cl}}$ when
$\lambda_{\rm{min}}$ is only barely lower than $\lambda_{\rm{lim}}$ and
$\lambda(\Delta M)$ quickly exceeds $\lambda_{\rm{lim}}$ for $\Delta M$
moving away from $\Delta M_{\rm{cl}}$.  To obtain a more
realistic measure of the actual accuracy of $\Delta M_{\rm{cl}}$ both, a
depth and a shape of the $\lambda$ minimum should be considered. A
deep, sharp minimum with $\lambda$ rising rapidly away from it means
that $\Delta M_{\rm{cl}}$ is well determined with a high degree of
confidence.  A broad, shallow $\lambda$ minimum means that the value of
$\Delta M_{\rm{cl}}$ is poorly constrained. 

Table~\ref{fit} shows that $\lambda_{\rm{min}}$ is reached for $\Delta M$ =
0 for all the investigated clusters, which means that only an upper limit
for the mass loss of each cluster can be determined.  In case of Pleiades
-- the youngest of the discussed clusters (see Table~\ref{mlr}), even the
distribution with $\Delta M = 0$ poorly describes the observed
distribution. This may be connected with an apparent deficit of massive
stars among the presently observed members (Moraux et
al. 2004). Fig.~\ref{fig2} shows the observed distributions of all four
clusters with the Salpeter function overplotted.  For NGC 6996 and Hyades
the distributions with $\Delta M = 0.1 M_{\sun}$ also describe
satisfactorily the observed distributions at the assumed significance level
so the upper limit for mass loss in these clusters is equal to 0.15
$M_{\sun}$. Note, however, that in case of Hyades the rise of
$\lambda(\Delta M)$ is so slow that a value of 0.20 $M_{\sun}$ only weakly
violates the 5 \% condition, so the limit of 0.25 $M_{\sun}$ is nearly
equally probable as 0.15 $M_{\sun}$. The same limit of 0.25 $M_{\sun}$ was
obtained from a comparison of the calculated distributions to observations
of 90 single Hyads with best known masses, carefully selected by de Bruijne
et al. (2001).  Finally, only the Salpeter function describes
satisfactorily the mass distribution in Praesepe and the rise of $\lambda
(\Delta M)$ is quite sharp (Table~\ref{fit}). { While the probability of
occuring $\lambda \ge 1.08$ is about 0.20, it drops to less than 0.01 for
$\lambda \ge 1.75$}.  We adopt a value of 0.05 $M{\sun}$ as an upper limit
for mass loss in this cluster.} Dividing the upper limit by the cluster age
we obtain an upper limit for an average MLR over the past cluster life. The
results (together with the adopted ages) are shown in Table~\ref{mlr}.

As we see, the derived MLRs in NGC 6996 and Hyades are not very
restrictive. However, the upper limit for mass loss and MLR in Praesepe
look interesting. The maximum mass loss is lower than the solar mass loss
of 0.07 $M_{\sun}$ determined by Sackmann \& Boothroyd (2003). The
maximum MLR is close to the theoretically estimated upper limit (see above)
and is lower than some of the empirical determinations (e.\thinspace
g. Mullan et al. 1992).

The result for Praesepe demonstrates a potential usefulness of the
suggested method. The presently attained accuracy of the method is limited
by two main factors: the accuracy of mass determination and scarcity of
data, { particularly for lower mass members}. The present mass
determinations extend only to $M \sim 0.8 M_{\sun}$. Deeper surveys,
including stars of spectral type K, i.\thinspace e. down
to 0.5 $M{\sun}$, are needed, together with an increased accuracy of mass
determination. { When masses of stars from the spectral range A-K in
stellar clusters with age $\sim$ 1 Gyr are known with relative accuracy of
0.01-0.02 $M_{\sun}$, it will be possible to detect average MLRs at the
level of $\sim 10^{-11} M_{\sun}$ or less. Such an accuracy should be
possible to reach with new generation instruments, like LSST, Gaia and
fibre-fed multiobjects spectrographs. The residual unresolved binaries will
add to the observational noise smearing the notch in the predicted mass
distribution (Fig.~\ref{fig1}). Fortunately, the Kolmogorov-Smirnov test is
only weakly sensitive to such an effect.

Systematic errors may be larger but this does not hurt -- such errors
influence only absolute mass determinations which are irrelevant
here. Moreover, they} can be offset by adequately varying the value of the
limiting mass for mass loss to occur.  If the future data are accurate
enough to detect a possible change of the slope of the mass distribution
with age it may be possible to analyze mass dependence of MLR. This,
however, needs a further increase of accuracy of mass determination {
and an extension of the observed mass distribution to still lower masses}.

\begin{table}
\caption[]{Upper limits for mass loss and
  the resulting upper limits for average MLRs of stars in the investigated
  clusters.}
\label{mlr}
\begin{tabular}{llcc} \hline

Cluster & Age &  Max. mass loss & Max. average MLR \\
        & Myr &  $M_{\sun}$ & $10^{-11} M_{\sun}$/yr \\
\hline
Pleiades & 100 & ? & ? \\
NGC 6996 & 325 & 0.15 & 43 \\
Hyades & 625 & 0.15-0.25 & 24-40  \\
Praesepe & 832 & 0.05 & 6 \\
\hline
\end{tabular}
\end{table}

\subsection{Conclusions}

A new method of mass loss determination in cool dwarfs is presented. It is
based on a comparison of the observed mass distribution in a stellar
cluster with a distribution modified by an assumed mass loss $\Delta M$
occurring during the cluster past life. The method was applied to four open
clusters with ages between 100 Myr and 1 Gyr. The most significant result
was obtained for Praesepe for which an upper limit of 0.05 $M_{\sun}$ for
$\Delta M$ was determined.  From the age of the cluster an upper limit for
average MLR can be calculated. It is equal to $6\times 10^{-11}
M_{\sun}$/yr. This is a promising result. With more { complete surveys
of cluster members and better mass determinations, a significantly better
estimate will be obtained}.

\begin{acknowledgements} 
We thank an anonymous referee for very helpful
 comments. This work was partly supported by the Ministry of Science and
 Higher Education grant 1 P03 016 28.
\end{acknowledgements}

\end{document}